\documentstyle[11pt]{article}
\bibliography{plain}
\title{\bf Lewenstein-Sanpera Decomposition for Iso-concurrence Decomposable
States } \vspace{20mm}
\author{
 S. J. Akhtarshenas$^{a,b,c}$
\thanks{E-mail:akhtarshenas@tabrizu.ac.ir}
 , M. A. Jafarizadeh$^{a,b,c}$ \thanks{E-mail:jafarizadeh@tabrizu.ac.ir}
\\
\\
$^a${\small Department of Theoretical Physics and Astrophysics,
Tabriz University, Tabriz 51664, Iran.} \\
$^b${\small Institute for Studies in Theoretical Physics and Mathematics,
 Tehran 19395-1795, Iran.} \\
$^c${\small Research Institute for Fundamental Sciences, Tabriz
51664, Iran.}} \pagebreak

\begin{document}
\maketitle \vspace{15mm}
\newpage
\begin{abstract}
We obtain Lewenstein-Sanpera decomposition of iso-concurrence
decomposable states of $2\otimes 2$ quantum systems. It is shown
that in these systems average concurrence of the decomposition is
equal to the concurrence of the state and also it is equal to the
amount of violation of positive partial transpose criterion. It is
also shown that the product states introduced by Wootters in [W.
K. Wootters, Phys. Rev. Lett. {\bf 80} 2245 (1998)] form the best
separable approximation ensemble for these states.


{\bf PACs Index: 03.65.Ud }
\end{abstract}
\pagebreak

\vspace{7cm}

\section{Introduction}
During the past decade an increasing study has been made on
entanglement, although it was discovered many decades ago by
Einstein and  Schr\"{o}dinger \cite{EPR,shcro}. This is because of
potential resource that entanglement provides for quantum
communication and information processing \cite{ben1,ben2,ben3}.
Entanglement  usually arises from quantum correlations between
separated subsystems which can not be created by local actions on
each subsystems. By definition, a bipartite mixed state $\rho$ is
said to be separable if it can be expressed as $$
\rho=\sum_{i}w_{i}\,\rho_i^{(1)}\otimes\rho_i^{(2)},\qquad w_i\geq
0, \quad \sum_{i}w_i=1, $$ where $\rho_i^{(1)}$ and $\rho_i^{(2)}$
denote density matrices of subsystems 1 and 2  respectively.
Otherwise the state is entangled.

The central tasks of quantum information theory is to characterize
and quantify entangled states. A first attempt in characterization
of entangled states has been made by Peres and Horodecki family
\cite{peres,horo}. It was shown that a necessary condition for
separability of a two partite system is that its partial
transposition be positive. Horodeckis showed that this condition
is sufficient for separability of composite systems only for
dimensions $2\otimes 2$ and $2 \otimes 3$.

There is also an increasing attention in quantifying entanglement,
particularly for mixed states of a bipartite system, and a number
of measures have been proposed \cite{ben3,ved1,ved2,woot}. Among
them the entanglement of formation has more importance, since it
intends to quantify the resources needed to create a given
entangled state.

An interesting description of entanglement is Lewenstein-Sanpera
decomposition \cite{LS}. Lewenstein and Sanpera in \cite{LS} have
shown that any two partite density matrix can be represented
optimally as a sum of a separable state (called best separable
approximation (BSA)) and an entangled state. They have also shown
that for 2-qubit systems the decomposition reduces to a mixture of
a mixed separable state and an entangled pure state, thus all
non-separability content of the state is concentrated in the pure
entangled state. This leads to an unambiguous measure of
entanglement for any 2-qubit state as entanglement of pure state
multiplied by the weight of pure part in the decomposition.

In the Ref. \cite{LS}, the numerical method for finding the BSA
has been reported. Also in $2\otimes2$ systems some analytical
results for special states were found in \cite{englert}. An
analytical expression for L-S decomposition of Bell decomposable
states and a class of states obtained from them via some LQCC
action is also obtained in \cite{jaf}.

In this paper we introduce iso-concurrence decomposable (ICD)
states. We give an analytical expression for L-S decomposition of
these states to show that the average concurrence of the
decomposition is equal to their concurrence in these states. As a
byproduct we also show that the concurrence of the considered
states is equal to amount of violation of positive partial
transpose criterion. It is also shown that the product states
introduced by Wootters in \cite{woot} form BSA ensemble for these
states.

Considering a real matrix parameterization of density matrices,
Verstraete et al. in \cite{vers1,vers2} showed that an local
quantum operations and classical communications (LQCC) corresponds
to left and right multiplication by a Lorentz matrix. They also
have shown that the real parameterization matrices
$R_{ij}=Tr(\rho\sigma_i\sigma_j)$ can be decomposed as $R=L_1\sum
L_2^{T}$, where $L_1$ and $L_2^{T}$ are proper Lorentz
transformations and $\sum$ has either a diagonal form correspond
to Bell decomposable states or it has a non-diagonal form. For the
case that the matrix $\sum$ becomes Bell decomposable state we
presented optimal decomposition for states obtained from them via
some LQCC actions \cite{jaf}. In other cases that $\sum$ has
non-diagonal form, corresponding matrix is a special cases of ICD
states, where  we consider it in this paper, and investigation of
action of LQCC on these states will appear in another paper.

The paper is organized as follows. In section 2 we introduce ICD
states, then we study their separability properties. The
concurrence of these states is evaluated in section 3, via the
method presented by Wootters in \cite{woot}. In section 4 we
obtain L-S decomposition of these states and we prove that thus
obtained decomposition is optimal. The paper is ended with a brief
conclusion in section 5.

\section{Iso-concurrence decomposable states}
In this section we define iso-concurrence decomposable (ICD)
states, then we give their separability condition. The
iso-concurrence states are defined by
\begin{eqnarray} \label{GBS1}
\left|\psi_1\right>=\cos{\theta}\left|\uparrow\uparrow\right>
+\sin{\theta}\left|\downarrow\downarrow\right>), \\
\label{GBS2}
\left|\psi_2\right>=\sin{\theta}\left|\uparrow\uparrow\right>
-\cos{\theta}\left|\downarrow\downarrow\right>), \\
\label{gBS3}
\left|\psi_3\right>=\cos{\theta}\left|\uparrow\downarrow\right>
+\sin{\theta}\left|\downarrow\uparrow\right>), \\
\label{GBS4}
\left|\psi_4\right>=\sin{\theta}\left|\uparrow\downarrow\right>
-\cos{\theta}\left| \downarrow\uparrow\right>).
\end{eqnarray}
It is quite easy to see that the above states are orthogonal and
thus they span the Hilbert space of $2\otimes 2$ systems. Also by
choosing $\theta=\frac{\pi}{4}$  the above states reduce to Bell
states. Now we can define ICD states as
\begin{equation} \label{ICDS}
\rho=\sum_{i=1}^{4}p_{i}\left|\psi_i\right>\left<\psi_i\right|,\quad\quad
0\leq p_i\leq 1,\quad \sum_{i=1}^{4}p_i=1.
\end{equation}
 These states form a four simplex (tetrahedral)  with its
vertices defined by $p_1=1$, $p_2=1$, $p_3=1$ and $p_4=1$,
respectively.

A necessary condition for separability of composite quantum
systems is presented by Peres \cite{peres}. He showed that if a
state is separable then the matrix obtained from partial
transposition must be positive. Horodecki family \cite{horo} have
shown that Peres criterion provides sufficient condition only for
separability of mixed quantum states of dimensions $2\otimes2$ and
$2\otimes3$. This implies that the state given in Eq. (\ref{ICDS})
is separable if and only if the following inequalities are
satisfied
\begin{eqnarray}
\label{ppt1}
(p_1-p_2)\leq \sqrt{4p_3p_4+(p_3-p_4)^2\sin^2{2\theta}}, \\
\label{ppt2}
(p_2-p_1)\leq \sqrt{4p_3p_4+(p_3-p_4)^2\sin^2{2\theta}}, \\
\label{ppt3}
(p_3-p_4)\leq \sqrt{4p_1p_2+(p_1-p_2)^2\sin^2{2\theta}}, \\
\label{ppt4} (p_4-p_3)\leq
\sqrt{4p_1p_2+(p_1-p_2)^2\sin^2{2\theta}}.
\end{eqnarray}

Inequalities (\ref{ppt1}) to (\ref{ppt4}) divide tetrahedral of
density matrices to five regions.  Central regions, defined by the
above inequalities, form a deformed octahedral and are separable
states. In four other regions one of the above inequality will not
hold, therefore they represent entangled states. In the next
sections we consider entangled states corresponding to violation
of inequality (\ref{ppt1}) i.e. the states which satisfy the
following inequality
\begin{equation}
\label{E1} (p_1-p_2)>\sqrt{4p_3p_4+(p_3-p_4)^2\sin^2{2\theta}}.
\end{equation}
All other ICD states can be obtain via local unitary
transformations.

\section{Concurrence}
In this section we first
review concurrence of mixed states. From the various measures
proposed to quantify entanglement, the entanglement of formation
has a special position which in fact intends to quantify the
resources needed to create a given entangled state \cite{ben3}.
Wootters in \cite{woot} has shown that for a 2-qubit system
entanglement of formation of a mixed state $\rho$ can be defined
as
\begin{equation}
E(\rho)=H\left(\frac{1}{2}+\frac{1}{2}\sqrt{1-C^2}\right),
\end{equation}
where $H(x)=-x\ln{x}-(1-x)\ln{(1-x)}$ is the binary entropy and
the concurrence $C(\rho)$ is defined by
\begin{equation}\label{concurrence}
C(\rho)=\max\{0,\lambda_1-\lambda_2-\lambda_3-\lambda_4\},
\end{equation}
where the $\lambda_i$ are the non-negative eigenvalues, with
$\lambda_1$ being the largest one, of the Hermitian matrix
$R\equiv\sqrt{\sqrt{\rho}{\tilde \rho}\sqrt{\rho}}$ and
\begin{equation}\label{rhotilde}
{\tilde \rho}
=(\sigma_y\otimes\sigma_y)\rho^{\ast}(\sigma_y\otimes\sigma_y),
\end{equation}
where $\rho^{\ast}$ is the complex conjugate of $\rho$ when it is
expressed in a fixed basis such as
$\{\left|\uparrow\uparrow\right>,
\left|\uparrow\downarrow\right>\},\{\left|\downarrow\uparrow\right>,
\left|\downarrow\downarrow\right>\}$ and $\sigma_y$ represent
Pauli matrix in local basis $\{\left|\uparrow\right>,
\left|\downarrow\right>\}$ .

In order to obtain the concurrence of ICD states we the follow
method presented by Wootters in \cite{woot}. Starting from
spectral decomposition for ICD states given in Eq. (\ref{ICDS}),
we define subnormalized orthogonal eigenvectors $\left|v_i\right>$
as
\begin{equation}\label{vvector}
\left|v_i\right>=\sqrt{p_i}\left|\psi_i\right>, \qquad
\left<v_i\mid v_j\right>=p_i \delta_{ij}.
\end{equation}
Now we can define states $\left|x_i\right>$ by
\begin{equation}\label{xvector}
\left|x_i\right>=\sum_{j}^{4}U_{ij}^{\ast}\left|v_i\right>, \qquad
\mbox{for}\quad i=1,2,3,4,
\end{equation}
such that
\begin{equation}\label{xortho}
\left<x_i\mid \tilde{x}_j\right>=(U\tau
U^T)_{ij}=\lambda_i\delta_{ij},
\end{equation}
where $\tau_{ij}=\left<v_i\mid v_j\right>$ is a symmetric but not
necessarily Hermitian matrix. To construct $\left|x_i\right>$ we
consider the fact that for any symmetric matrix $\tau$ one can
always find a unitary matrix $U$ in such a way that $\lambda_i$
are real and non-negative, that is, they are the square roots of
eigenvalues of $\tau\tau^{\ast}$ which are same as eigenvalues of
$R$. Moreover one can always find $U$ such that $\lambda_1$ being
the largest one.

By using the above protocol we get the following expression for
state of $\rho$ given in Eq. (\ref{ICDS})
\begin{equation}\label{tau}
\tau=\left(\begin{array}{cccc}
-p_1\sin{2\theta} & \sqrt{p_1 p_2}\cos{2\theta} & 0 & 0 \\
\sqrt{p_1 p_2} \cos{2\theta} & p_2\sin{2\theta} & 0 & 0 \\
0 & 0 & p_3\sin{2\theta} & -\sqrt{p_3 p_4}\cos{2\theta} \\
0 & 0 & -\sqrt{p_3 p_4} \cos{2\theta} & -p_4\sin{2\theta}
\end{array}\right).
\end{equation}
Now it is easy to evaluate $\lambda_i$ which yields
\begin{equation}\label{lambda1234}
\begin{array}{c}
\lambda_1=\frac{1}{2}\left((p_1-p_2)\sin{2\theta}
+\sqrt{4p_1p_2+(p_1-p_2)^2\sin^2{2\theta}}\right), \\
\lambda_2=\frac{1}{2}\left((p_2-p_1)\sin{2\theta}
+\sqrt{4p_1p_2+(p_1-p_2)^2\sin^2{2\theta}}\right), \\
\lambda_3=\frac{1}{2}\left((p_3-p_4)\sin{2\theta}
+\sqrt{4p_3p_4+(p_3-p_4)^2\sin^2{2\theta}}\right), \\
\lambda_4=\frac{1}{2}\left((p_4-p_3)\sin{2\theta}
+\sqrt{4p_3p_4+(p_3-p_4)^2\sin^2{2\theta}}\right).
\end{array}
\end{equation}
Thus one can evaluate the concurrence of ICD states as
\begin{equation}\label{CICDS}
C=(p_1-p_2)\sin{2\theta}-\sqrt{4p_3p_4+(p_3-p_4)^2\sin^2{2\theta}}.
\end{equation}
It is worth to note that thus obtained concurrence is equal to the
amount of violation of inequality (\ref{E1}).

Finally we introduce the unitary matrix $U$ which is going to be
used later
\begin{equation}\label{U}
U=\left(\begin{array}{cccc}
i\alpha_1 & -i\alpha_2 & 0 & 0 \\
\alpha_2 & \alpha_1 & 0 & 0 \\
0 & 0 & \alpha_3 & -\alpha_4 \\
0 & 0 & i\alpha_4 & i\alpha_3
\end{array}\right),
\end{equation}
where
$$
\alpha_1=\frac{\left((p_1+p_2)\sin{2\theta}+
\sqrt{4p_1p_2+(p_1-p_2)^2\sin^2{2\theta}}\right)}
{\sqrt{2}\left(4p_1p_2\cos^2{2\theta}+(p_1+p_2)^2\sin^2{2\theta}+
(p_1+p_2)\sin{2\theta}\sqrt{4p_1p_2+(p_1-p_2)^2\sin^2{2\theta}}\right)},
$$
$$
\alpha_2=\frac{\sqrt{2p_1p_2}\cos{2\theta}}
{\left(4p_1p_2\cos^2{2\theta}+(p_1+p_2)^2\sin^2{2\theta}+
(p_1+p_2)\sin{2\theta}\sqrt{4p_1p_2+(p_1-p_2)^2\sin^2{2\theta}}\right)},
$$
\begin{equation}\label{alpha}
\end{equation}
$$
\alpha_3=\frac{\left((p_3+p_4)\sin{2\theta}+
\sqrt{4p_3p_4+(p_3-p_4)^2\sin^2{2\theta}}\right)}
{\sqrt{2}\left(4p_3p_4\cos^2{2\theta}+(p_3+p_4)^2\sin^2{2\theta}+
(p_3+p_4)\sin{2\theta}\sqrt{4p_3p_4+(p_3-p_4)^2\sin^2{2\theta}}\right)},
$$
$$
\alpha_4=\frac{\sqrt{2p_3p_4}\cos{2\theta}}
{\left(4p_3p_4\cos^2{2\theta}+(p_3+p_4)^2\sin^2{2\theta}+
(p_3+p_4)\sin{2\theta}\sqrt{4p_3p_4+(p_3-p_4)^2\sin^2{2\theta}}\right)}.
$$

\section{Lewenstein-Sanpera decomposition}
According to Lewenstein-Sanpera decomposition \cite{LS}, any
2-qubit  density matrix $\rho$ can be written as
\begin{equation}\label{LSD}
\rho=\lambda\rho_{sep}+(1-\lambda)\left|\psi\right>\left<\psi\right|,
\quad\quad
\lambda\in[0,1],
\end{equation}
where $\rho_{sep}$ is a separable density matrix and
$\left|\psi\right>$ is a pure entangled state. The
Lewenstein-Sanpera decomposition of a given density matrix $\rho$
is not unique and, in general, there is a continuum set of L-S
decomposition to choose from. The optimal decomposition is,
however, unique for which $\lambda$ is maximal and
\begin{equation}\label{LSDopt}
\rho=\lambda^{(opt)}\rho_{sep}^{(opt)}
+(1-\lambda^{(opt)})|\psi^{(opt)}\left>\right<\psi^{(opt)}|\;,
\quad\quad \lambda^{(opt)}\in[0,1].
\end{equation}
 Lewenstein and Sanpera in \cite{LS} have shown that any other
  decomposition of the form
 $\rho={\tilde \lambda}{\tilde \rho}_{sep}
 +(1-{\tilde \lambda})|{\tilde \psi}\left>\right<{\tilde \psi}|$
 with ${\tilde
 \rho}\neq\rho^{(opt)}$ necessarily implies that ${\tilde
 \lambda}<\lambda^{(opt)}$ \cite{LS}.
We refer to Eq. (\ref{LSDopt}) as optimal decomposition in the
sense that $\lambda$ is maximal and $\rho_s$ is the best separable
approximation (BSA).

 Here in this section we would like to obtain L-S decomposition
for ICD states. Let us consider entangled state $\rho$ belonging
to entangled region defined by Eq. (\ref{E1}). We start by writing
density matrix $\rho$ as a convex sum of pure state
$\left|\psi_1\right>$ and separable state $\rho_s$ as
\begin{equation}\label{LSD1}
\rho=\lambda\rho_s+(1-\lambda)\left|\psi_1\right>\left<\psi_1\right|.
\end{equation}
where, $\rho_s$ corresponds to intersection of the line passing
through the points corresponding to
$\left|\psi_1\right>\left<\psi_1\right|$ and $\rho$ with
2-dimensional manifold defined by the equality in relation
(\ref{ppt1}). Expanding separable state $\rho_s$ as
$\rho_s=\sum_{i=1}^{4}p_i^{\prime}\left|\psi_i\right>\left<\psi_i\right|$
and using Eq. (\ref{ICDS}) for $\rho$ we arrive at the following
results
\begin{equation}\label{pprime}
p_1^{\prime}=\frac{p_1+\lambda-1}{\lambda}, \qquad
p_i^{\prime}=\frac{p_i}{\lambda} \quad \mbox{for}\quad i=2,3,4,
\end{equation}
where
\begin{equation}\label{lambda}
\lambda=1-p_1+p_2+\sqrt{\frac{4p_3p_4}{\sin^2{2\theta}}+(p_3-p_4)^2}.
\end{equation}
Now using the fact that for pure states defined in Eqs.
(\ref{GBS1}) to (\ref{GBS4}) we have
$C(\left|\psi_i\right>)=\sin{2\theta}$, one can easily evaluate
average concurrence of the decomposition given in Eq. (\ref{LSD1})
as
\begin{equation}\label{avecon}
(1-\lambda)C(\left|\psi_1\right>)=
(p_1-p_2)\sin{2\theta}-\sqrt{4p_3p_4+(p_3-p_4)^2\sin^2{2\theta}}.
\end{equation}
Equation (\ref{avecon}) shows that average concurrence of $\rho$
in the L-S decomposition given in (\ref{LSD1}) is equal to its
concurrence. As the concurrence of a mixed state is defined as the
minimum of the average concurrence over all decompositions of the
state in terms of pure states \cite{woot}, so according to Eq.
(\ref{LSD1}) we should have $C(\rho)\le (1-\lambda)C(\psi)$. But
Eq. (\ref{avecon}) shows that the optimal decomposition of ICD
states saturate this inequality.

In the rest of this section we will prove that the decomposition
(\ref{LSD1}) is the optimal one. To do so we have to find a
decomposition for $\rho_s$ in terms of product states
$\left|e_\alpha,f_\alpha\right>$, i.e.
\begin{equation}
\rho_s=\sum_{\alpha}\Lambda_\alpha
\left|e_\alpha,f_\alpha\right>\left<e_\alpha,f_\alpha\right|
\end{equation}
such that the following conditions are satisfied \cite{LS}

i) All $\Lambda_\alpha$ are maximal with respect to
$\rho_\alpha=\Lambda_\alpha
\left|e_\alpha,f_\alpha\right>\left<e_\alpha,f_\alpha\right|
+(1-\lambda)\left|\psi_1\right>\left<\psi_1\right|$ and projector
$P_\alpha=\left|e_\alpha,f_\alpha\right>\left<e_\alpha,f_\alpha\right|$.

ii) All pairs $(\Lambda_\alpha,\Lambda_\beta)$ are maximal with
respect to $\rho_{\alpha\beta}=\Lambda_\alpha
\left|e_\alpha,f_\alpha\right>\left<e_\alpha,f_\alpha\right|
+\Lambda_\beta
\left|e_\beta,f_\beta\right>\left<e_\beta,f_\beta\right|
+(1-\lambda)\left|\psi_1\right>\left<\psi_1\right|$ and the pairs
of projector $(P_\alpha,P_\beta)$.

Then according to \cite{LS} $\rho_s$ is BSA and the decomposition
given in Eq. (\ref{LSD1}) is optimal.

Lewenstein and Sanpera in \cite{LS} have shown that $\Lambda$ is
maximal with respect to $\rho$ and
$P=\left|\psi\right>\left<\psi\right|$ iff a) if
$\left|\psi\right>\not\in {\cal R}(\rho)$ then $\Lambda=0$, and b)
if $\left|\psi\right>\in {\cal R}(\rho)$ then
$\Lambda=\left<\psi\right|\rho^{-1}\left|\psi\right>^{-1}>0$. They
have also shown that a pair $(\Lambda_1,\Lambda_2)$ is maximal
with respect to $\rho$ and a pair of projectors $(P_1,P_2)$ iff:
a) if $\left|\psi_1\right>$, $\left|\psi_2\right>$ do not belong
to ${\cal R}(\rho)$ then $\Lambda_1=\Lambda_2=0$; b) if
$\left|\psi_1\right>$ does not belong, while
$\left|\psi_2\right>\in{\cal R}(\rho)$ then $\Lambda_1=0$,
$\Lambda_2=\left<\psi_2\right|\rho^{-1}\left|\psi_2\right>^{-1}$;
c) if $\left|\psi_1\right>$, $\left|\psi_2\right>\in {\cal
R}(\rho)$ and $\left<\psi_1\right|\rho^{-1}\left|\psi_2\right>=0$
then
$\Lambda_i=\left<\psi_i\right|\rho^{-1}\left|\psi_i\right>^{-1}$,
$i=1,2$; d) finally, if $\left|\psi_1\right>,
\left|\psi_2\right>\in {\cal R}(\rho)$ and
$\left<\psi_1\right|\rho^{-1}\left|\psi_2\right>\neq 0$ then

\begin{equation}\label{Lambda12}
\begin{array}{lr}
\Lambda_1= &(\left<\psi_2\right|\rho^{-1}\left|\psi_2\right>
-\mid\left<\psi_1\right|\rho^{-1}\left|\psi_2\right>\mid)/D, \\
\Lambda_2= &(\left<\psi_1\right|\rho^{-1}\left|\psi_1\right>
-\mid\left<\psi_1\right|\rho^{-1}\left|\psi_2\right>\mid)/D,
\end{array}
\end{equation}
where
$D=\left<\psi_1\right|\rho^{-1}\left|\psi_1\right>\left<\psi_2\right|
\rho^{-1}\left|\psi_2\right>
-\mid\left<\psi_1\right|\rho_{-1}\left|\psi_2\right>\mid^2$.

Now we return to show that the decomposition given in (\ref{LSD1})
is optimal. Wootters in \cite{woot} has shown that any $2\otimes
2$ separable density matrix can be expanded in terms of following
product states
\begin{equation}\label{z1234-1}
\left|z_1\right>=\frac{1}{2}\left(e^{i\theta_1}\left|x_1\right>
+e^{i\theta_2}\left|x_2\right>+e^{i\theta_3}\left|x_3\right>
+e^{i\theta_4}\left|x_4\right>\right),
\end{equation}
\begin{equation}\label{z1234-1}
\left|z_2\right>=\frac{1}{2}\left(e^{i\theta_1}\left|x_1\right>
+e^{i\theta_2}\left|x_2\right>-e^{i\theta_3}\left|x_3\right>
-e^{i\theta_4}\left|x_4\right>\right),
\end{equation}
\begin{equation}\label{z1234-1}
\left|z_3\right>=\frac{1}{2}\left(e^{i\theta_1}\left|x_1\right>
-e^{i\theta_2}\left|x_2\right>+e^{i\theta_3}\left|x_3\right>
-e^{i\theta_4}\left|x_4\right>\right),
\end{equation}
\begin{equation}\label{z1234-1}
\left|z_4\right>=\frac{1}{2}\left(e^{i\theta_1}\left|x_1\right>
-e^{i\theta_2}\left|x_2\right>-e^{i\theta_3}\left|x_3\right>
+e^{i\theta_4}\left|x_4\right>\right),
\end{equation}
provided that $\lambda_1-\lambda_2-\lambda_3-\lambda_4\leq 0$.
Now, the zero concurrence is guaranteed by choosing phases
$\theta_i$, $i=1,2,3,4$ satisfying the relation
$\sum_{j=1}e^{2i\theta_j}\lambda_j=0$.

Now using the fact that for marginal  separable states $\rho_s$
(located at the boundary of separable region)  the eigenvalues
$\lambda_i$ satisfy the constraint
$\lambda_1-\lambda_2-\lambda_3-\lambda_4=0$, we can choose the
phase factors $\theta_i$ as
$\theta_2=\theta_3=\theta_4=\theta_1+\frac{\pi}{2}$. Choosing
$\theta_1=0$  we arrive at the following product ensemble for
$\rho_s$
\begin{equation}
\begin{array}{rl}
\left|z_1\right>=\frac{1}{2}
(-i(\alpha_1^{\prime}+\alpha_2^{\prime})\sqrt{p_1^{\prime}}\left|\psi_1\right>
-i(\alpha_1^{\prime}-\alpha_2^{\prime})\sqrt{p_2^{\prime}}\left|\psi_2\right>
\\
-(\alpha_4^{\prime}+i\alpha_3^{\prime})\sqrt{p_3^{\prime}}\left|\psi_3\right>
-(\alpha_3^{\prime}-i\alpha_4^{\prime})\sqrt{p_4^{\prime}}\left|\psi_4\right>),
\\
\left|z_2\right>=\frac{1}{2}
(-i(\alpha_1^{\prime}+\alpha_2^{\prime})\sqrt{p_1^{\prime}}\left|\psi_1\right>
-i(\alpha_1^{\prime}-\alpha_2^{\prime})\sqrt{p_2^{\prime}}\left|\psi_2\right>
\\
+(\alpha_4^{\prime}+i\alpha_3^{\prime})\sqrt{p_3^{\prime}}\left|\psi_3\right>
+(\alpha_3^{\prime}-i\alpha_4^{\prime})\sqrt{p_4^{\prime}}\left|\psi_4\right>),
\\
\left|z_3\right>=\frac{1}{2}
(-i(\alpha_1^{\prime}-\alpha_2^{\prime})\sqrt{p_1^{\prime}}\left|\psi_1\right>
+i(\alpha_1^{\prime}+\alpha_2^{\prime})\sqrt{p_2^{\prime}}\left|\psi_2\right>
\\
+(\alpha_4^{\prime}-i\alpha_3^{\prime})\sqrt{p_3^{\prime}}\left|\psi_3\right>
+(\alpha_3^{\prime}+i\alpha_4^{\prime})\sqrt{p_4^{\prime}}\left|\psi_4\right>),
\\
\left|z_4\right>=\frac{1}{2}
(-i(\alpha_1^{\prime}-\alpha_2^{\prime})\sqrt{p_1^{\prime}}\left|\psi_1\right>
+i(\alpha_1^{\prime}+\alpha_2^{\prime})\sqrt{p_2^{\prime}}\left|\psi_2\right>
\\
-(\alpha_4^{\prime}-i\alpha_3^{\prime})\sqrt{p_3^{\prime}}\left|\psi_3\right>
-(\alpha_3^{\prime}+i\alpha_4^{\prime})\sqrt{p_4^{\prime}}\left|\psi_4\right>),
\end{array}
\end{equation}
where $\alpha_i^{\prime}$ are defined in the same form as given in
Eq. (\ref{alpha}) provided that they are expressed in terms of
$p_i^{\prime}$ (coordinates of $\rho_s$).

Now in order to prove that the decomposition given in Eq.
(\ref{LSD}) is optimal we first consider the cases that $\rho$ has
full rank. Let us consider the set of four product vectors
$\{\left|z_i\right>\}$ and one entangled state
$\left|\psi_1\right>$. In Ref. \cite{woot} it is shown that the
ensemble $\{\left|z_i\right>\}$ are linearly independent, also it
is straightforward to see that three vectors
$\left|z_\alpha\right>$ , $\left|z_\beta\right>$ and
$\left|\psi_1\right>$ are linearly independent. Now let us
consider matrices $\rho_\alpha=\Lambda_\alpha \left|
z_\alpha\right>\left<z_\alpha\right|
+(1-\lambda)\left|\psi_1\right>\left<\psi_1\right|$. Due to
independence of $\left|z_\alpha\right>$ and $\left|\psi_1\right>$
we can deduce that the range of $\rho_\alpha$ is two dimensional,
thus after restriction to its range and defining dual basis
$\left|{\hat z}_\alpha\right>$ and $\left|{\hat \psi}_1\right>$ we
can expand restricted inverse $\rho_\alpha^{-1}$ as
$\rho_{\alpha}^{-1}=\Lambda_\alpha^{-1}|{\hat
z}_\alpha\left>\right<{\hat z}_\alpha| +(1-\lambda)^{-1}|{\hat
\psi}_1\left>\right<{\hat \psi}_1|$ (see appendix). Using Eq.
(\ref{A1}) it is easy to see that
$\left<z_\alpha|\rho_{\alpha}^{-1}
|z_\alpha\right>=\Lambda_\alpha^{-1}$. This shows that
$\Lambda_\alpha$ are maximal with respect to $\rho_\alpha$ and the
projector $P_\alpha=\left|z_\alpha\right>\left<z_\alpha\right|$.

Similarly considering matrices $\rho_{\alpha\beta}=\Lambda_\alpha
\left|z_\alpha\right>\left<z_\alpha\right| +\Lambda_\beta
\left|z_\beta\right>\left<z_\beta\right|
+(1-\lambda)\left|\psi_1\right>\left<\psi_1\right|$ and
considering the independence of  vectors $\left|z_\alpha\right>$,
$\left|z_\beta\right>$ and $\left|\psi_1\right>$ we see that rang
of $\rho_{\alpha\beta}$ is three dimensional where after
restriction to its range and defining their dual basis
$\left|{\hat z}_\alpha\right>$, $\left|{\hat z}_\beta\right>$ and
$\left|{\hat \psi}_1\right>$ we can write restricted inverse
$\rho_{\alpha\beta}^{-1}$ as
$\rho_{\alpha\beta}^{-1}=\Lambda_\alpha^{-1}|{\hat
z}_\alpha\left>\right<{\hat z}_\alpha|+\Lambda_\beta^{-1}|{\hat
z}_\beta\left>\right<{\hat z}_\beta|+(1-\lambda)^{-1}|{\hat
\psi}_1\left>\right<{\hat \psi}_1|$. Then it is straightforward to
get $\left<{\hat
e}_\alpha\right|\rho_{\alpha\beta}^{-1}\left|{\hat
z}_\alpha\right>=\Lambda_\alpha^{-1}$, $\left<{\hat
z}_\beta\right|\rho_{\alpha\beta}^{-1}\left|{\hat
z}_\beta\right>=\Lambda_\beta^{-1}$ and $\left<{\hat
z}_\alpha\right|\rho_{\alpha\beta}^{-1}\left|{\hat
z}_\beta\right>=0$. This implies that the pair
$(\Lambda_\alpha,\Lambda_\beta)$ are maximal with respect to
$\rho_{\alpha\beta}$ and the pair of projectors
$(P_\alpha,P_\beta)$, thus complete the proof that the
decomposition given in Eq. (\ref{LSD}) is optimal.

We now consider the cases that $\rho$ has not full rank. First let
us consider $p_2^{\prime}=0$ case. In this case the pairs
$\{\left|z_1\right>,\left|z_2\right>\}$ and also
$\{\left|z_3\right>,\left|z_4\right>\}$ are no longer independent
with respect to $\left|\psi_1\right>$. In this cases we can easily
evaluate $\left|\psi_1\right>$ in terms of $\left|z_1\right>$ and
$\left|z_2\right>$ then matrix $\rho_{12}$ can be written in terms
of two basis $\left|z_1\right>$ and $\left|z_2\right>$ which
yields after some calculations,
$\left<z_1|\rho_{12}^{-1}|z_1\right>=
\frac{\Lambda_2(\alpha_1^{\prime}+\alpha_2^{\prime})^2p_1^{\prime}
+(1-\lambda)}{\Gamma_{12}}$, $\left<z_2|\rho_{12}^{-1}|z_2\right>=
\frac{\Lambda_1(\alpha_1^{\prime}+\alpha_2^{\prime})^2p_1^{\prime}
+(1-\lambda)}{\Gamma_{12}}$ and
$\left<z_1|\rho_{12}^{-1}|z_2\right>=
\frac{-(1-\lambda)}{\Gamma_{12}}$ where $\Gamma_{12}=
\Lambda_1\Lambda_2(\alpha_1^{\prime}+\alpha_2^{\prime})^2p_1^{\prime}
+(1-\lambda)(\Lambda_1+\Lambda_2)$. Using the above results
together with Eqs. (\ref{Lambda12}) we obtain the maximality of
pair $(\Lambda_1,\Lambda_{2})$ with respect to $\rho_{12}$ and the
pair of projectors $(P_1,P_{2})$. Similarly we can express
$\left|\psi_1\right>$ in terms of $\left|z_3\right>$ and
$\left|z_4\right>$ and evaluate $\rho_{34}^{-1}$, which get
$\left<z_3|\rho_{34}^{-1}|z_3\right>=
\frac{\Lambda_4(\alpha_1^{\prime}-\alpha_2^{\prime})^2p_1^{\prime}
+(1-\lambda)}{\Gamma_{34}}$, $\left<z_4|\rho_{34}^{-1}|z_4\right>=
\frac{\Lambda_3(\alpha_1^{\prime}-\alpha_2^{\prime})^2p_1^{\prime}
+(1-\lambda)}{\Gamma_{34}}$ and
$\left<z_3|\rho_{34}^{-1}|z_4\right>=
\frac{-(1-\lambda)}{\Gamma_{34}}$ with  $\Gamma_{34}=
\Lambda_3\Lambda_4(\alpha_1^{\prime}-\alpha_2^{\prime})^2p_1^{\prime}
+(1-\lambda)(\Lambda_3+\Lambda_4)$, together with Eqs.
(\ref{Lambda12}) we obtain the maximality of pair
$(\Lambda_3,\Lambda_{4})$ with respect to $\rho_{34}$ and the pair
of projectors $(P_3,P_{4})$. For other choices of $\alpha$ and
$\beta$ three vectors $\left|z_\alpha\right>$
,$\left|z_\beta\right>$ and $\left|\psi_1\right>$ remain linearly
independent, thus with the above mentioned method we can prove the
maximality of pairs $(\Lambda_\alpha,\Lambda_\beta)$.

Also in the cases of $p_3^{\prime}=0$ or $p_4^{\prime}=0$, three
vectors $\left|z_\alpha\right>$ ,$\left|z_\beta\right>$ and
$\left|\psi_1\right>$ remain linearly independent thus maximality
of pairs $(\Lambda_\alpha,\Lambda_\beta)$ can be proven in the
same method of full rank one.

Now let us consider the cases that rank $\rho$ is 2. In this case
we have to write the separable part in terms of rank one
projection operators which are not product any more  but as far as
the maximality of pairs(the requirement for the maximality of
separable part of density matrix in this case) are concerned the
separability of the projectors does not play any role.

For $p_2=p_3=0$ we have $p_1^{\prime}=p_4^{\prime}=\frac{1}{2}$,
therefore $\rho_s$ can be written as $$
\rho_s=\frac{1}{2}\left(\left|\psi_1\right>\left<\psi_1\right|+
\left|\psi_4\right>\left<\psi_4\right|\right)
=\frac{1}{2}\left(\left|e_1,f_1\right>\left<e_1,f_1\right|+
\left|e_2,f_2\right>\left<e_2,f_2\right|\right), $$ where
$\left|e_{1,2},f_{1,2}\right>=\left(\cos{\theta}\left|\uparrow\right>
\mp
i\sin{\theta}\left|\downarrow\right>\right)\otimes\frac{1}{\sqrt{2}}
\left(\left|\uparrow\right>\pm i\left|\downarrow\right>\right)$.
Now by writing $\left|\psi_1\right>$ in terms of product states
$\left|e_{1},f_{1}\right>$ and $\left|e_{2},f_{2}\right>$ the
matrix $\rho_{12}$ can be expressed in terms of two basis
$\left|e_1,f_1\right>$, $\left|e_2,f_2\right>$ and we get
$\left<e_1,f_1|\rho_{12}^{-1}|e_1,f_1\right>=
\frac{2\Lambda_2+(1-\lambda)}{\Gamma_{12}}$,
$\left<e_2,f_2|\rho_{12}^{-1}|e_2,f_2\right>= \frac{2\Lambda_1
+(1-\lambda)}{\Gamma_{12}}$ and
$\left<e_1,f_1|\rho_{12}^{-1}|e_2,f_2\right>=
\frac{-(1-\lambda)}{\Gamma_{12}}$ where $\Gamma_{12}=
2\Lambda_1\Lambda_2+(1-\lambda)(\Lambda_1+\Lambda_2)$. Using the
above results together with Eqs. (\ref{Lambda12}) we obtain the
maximality of pair $(\Lambda_1,\Lambda_{2})$ with respect to
$\rho_{12}$ and the pair of projectors $(P_1,P_{2})$. The other
rank-2 cases can be proved via similar method.

\section{Conclusion}
We have derived Lewenstein-Sanpera decomposition for
iso-concurrence decomposable states. It is shown that for these
states the average concurrence of L-S decomposition is equal to
the concurrence of the states. It is also shown that Wootters
product states defined in \cite{woot} form the best separable
approximation ensemble for these states.

{\large \bf Appendix }

Let us consider the set of linearly independent vectors
$\{\left|\phi_i\right>\}$, then one can define their dual vectors
$\{\left|{\hat \phi}_i\right>\}$ such that the following relation
\begin{equation}\label{A1}
\left<{\hat \phi}_i\mid\phi_j\right>=\delta_{ij}
\end{equation}
hold. It is straightforward to show that the
$\{\left|\phi_i\right>\}$ and their dual $\{\left|{\hat
\phi}_i\right>\}$ posses the following completeness relation
\begin{equation}\label{ids}
\sum_{i}|{\hat \phi}_i\left>\right<\phi_i|=I,
\qquad\sum_{i}|\phi_i\left>\right<{\hat \phi}_i|=I.
\end{equation}
Consider an invertible     operator $M$ which is expanded in terms
of states $\left|\phi_i\right>$ as
\begin{equation}\label{M}
M=\sum_{i}a_{ij}\left|\phi_i\right>\left<\phi_j\right|
\end{equation}
Then the inverse of $M$ denoted by $M^{-1}$ can be expanded in
terms of dual bases as
\begin{equation}\label{Minverse}
M^{-1}=\sum_{i}b_{ij}|{\hat \phi}_i\left>\right<{\hat \phi}_j|
\end{equation}
where $b_{ij}=(A^{-1})_{ij}$ and $A_{ij}=a_{ij}$.


\begin{thebibliography}{99}
\bibitem{EPR}{\sc A. Einstein, B. Podolsky and Rosen, }
{\em  Phys. Rev. {\bf 47}, 777 (1935).}
\bibitem{shcro}{\sc E. Schr\"{O}dinger, }
{\em Naturwissenschaften. {\bf 23} 807 (1935).}
\bibitem{ben1}{\sc C. H. Bennett, and S. J. Wiesner,}
{\em Phys. Rev. Lett. {\bf 69}, 2881 (1992).}
\bibitem{ben2}{\sc C. H. Bennett, G. Brassard,
C. Cr\'{e}peau, R. jozsa, A Peres and W. K. Wootters,} {\em Phys.
Rev. Lett. {\bf 70}, 1895 (1993).}
\bibitem{ben3}{\sc C. H. Bennett, D. P. DiVincenzo, J. A. Smolin and W.K.
Wootters,} {\em Phys. Rev. A {\bf 54}, 3824 (1996).}
\bibitem{peres}{\sc A. Peres, }{\em Phys. Rev. Lett. {\bf 77} 1413 (1996).}
\bibitem{horo}{\sc M. Horodecki, P. Horodecki and R. Horodecki, }
{\em Phys. Lett. A  {\bf 223} 1 (1996).}
\bibitem{ved1}{\sc V. Vedral, M. B. Plenio, M. A. Rippin and P. L. Knight,}
 {\em Phys. Rev. Lett. {\bf 78}, 2275 (1995).}
\bibitem{ved2}{\sc V. Vedral and M. B. Plenio,}
 {\em Phys. Rev. A {\bf 57}, 1619 (1998).}
\bibitem{woot}{\sc W. K. Wootters, }
{\em Phys. Rev. Lett. {\bf 80} 2245 (1998).}
\bibitem{LS}{\sc M. Lewenstein and A. Sanpera, }
 { \em Phys. Rev. Lett. {\bf 80,} 2261 (1998). }
\bibitem{englert}{\sc B.-G Englert and N. Metwally, }
 { \em J. Mod. Opt. {\bf 47,} 2221 (2000). }
\bibitem{jaf}{\sc S. J. Akhtarshenas and M. A. Jafarizadeh, }
 { \em eprint quant-ph/0207161 {\bf }  (2002). }
\bibitem{vers1}{\sc F. Verstraete, J. Dehaene and B. DeMoor, }
 {\em Phys. Rev. A {\bf 64}, 010101 (2001).}
\bibitem{vers2}{\sc F. Verstraete, J. Dehaene and B. DeMoor, }
 {\em Phys. Rev. A {\bf 65}, 032308 (2002).}
\end{thebibliography}
\end{document}